\documentclass{iaus}
\usepackage{graphicx}

\title[Pair-Instability Supernovae] %% give here short title %%
{Pair-Instability Explosions: \\ observational evidence}

\author[Avishay Gal-Yam]   %% give here short author list %%
{Avishay Gal-Yam}

\affiliation{Department of Particle Physics and Astrophysics, \\ Weizmann Institute of Science,
76100 Rehovot, Israel \\ email: {\tt avishay.gal-yam@weizmann.ac.il}}

\pubyear{2012}
\volume{279}  %% insert here IAU Symposium No.
\pagerange{xxx--yyy}
\jname{Death of Massive Stars: Supernovae and Gamma-Ray Bursts}
\editors{P. Roming, N. Kawai \& E. Pian, eds.}
\begin{document}

\maketitle

\begin{abstract}
It has been theoretically predicted many decades ago that extremely massive stars 
that develop large oxygen cores will become dynamically unstable, due to electron-positron 
pair production. The collapse of such oxygen cores leads to powerful 
thermonuclear explosions that unbind the star and can produce, in some cases, many solar masses
of radioactive $^{56}$Ni. For many years, no examples of this process were observed in nature. 
Here, I briefly review recent observations of luminous supernovae that likely result from
pair-instability explosions, in the nearby and distant Universe. 
 
%\keywords{Keyword1, keyword2, keyword3, etc.}
\end{abstract}

\firstsection % if your document starts with a section,
\section{Introduction}

The pair-instability explosion mechanism (e.g., \cite[Rakavy \& Shaviv 1967]{RS67};
\cite[Barkat, Rakavy \& Sack 1967]{BRS67}; \cite[Bond et al. 1984]{B+84}; \cite[Heger \& Woosley 2002]{HW02}; 
\cite[Scannapieco et al. 2005]{S+05}; \cite[Waldman 2008]{W08}) was predicted to
occur during the evolution of very massive stars that develop oxygen cores above a critical mass threshold ($\sim50$\,M$_{\odot}$).
These cores achieve high temperatures at relatively low densities (e.g., Fig. 1 of \cite[Waldman 2008]{W08}). 
Significant amounts of electron-positron pairs are created prior
to oxygen ignition; loss of pressure support, rapid contraction, and explosive oxygen ignition follow, leading to a powerful explosion that
disrupts the star. Extensive theoretical work indicates this result is unavoidable for massive oxygen cores; when the core mass in 
question is large enough ($\sim100$\,M$_{\odot}$; e.g., \cite[Heger \& Woosley 2002]{HW02}; \cite[Waldman 2008]{W08}) 
many solar masses of radioactive nickel are naturally produced. Such Ni-rich events will be extremely luminous and therefore easy
to observe. On the other hand, oxygen cores that are massive enough to become pair unstable were predicted to evolve,
according to most stellar-evolution models, only in stars of exceedingly large initial masses (many hundreds times M$_{\odot}$; e.g.,
\cite[Yoshida \& Umeda 2010]{YU10}; though see \cite[Langer et al. 2007]{L+07}), unless
the stars are assumed to have very low initial metallicity. For this reason it was often assumed that pair-instability supernovae
(PISN) only occurred among population III stars at very high redshifts. Recently, we have shown that luminous events that match
the predictions of PISN models very well (SLSN-R; \cite[Gal-Yam 2012]{G12}) do occur in dwarf galaxies in the local 
Universe (e.g., SN 2007bi \cite[Gal-Yam et al. 2009]{G+09}). 

\section{Early candidates}

\begin{figure}[h]
\centering
\includegraphics[width=1\textwidth]{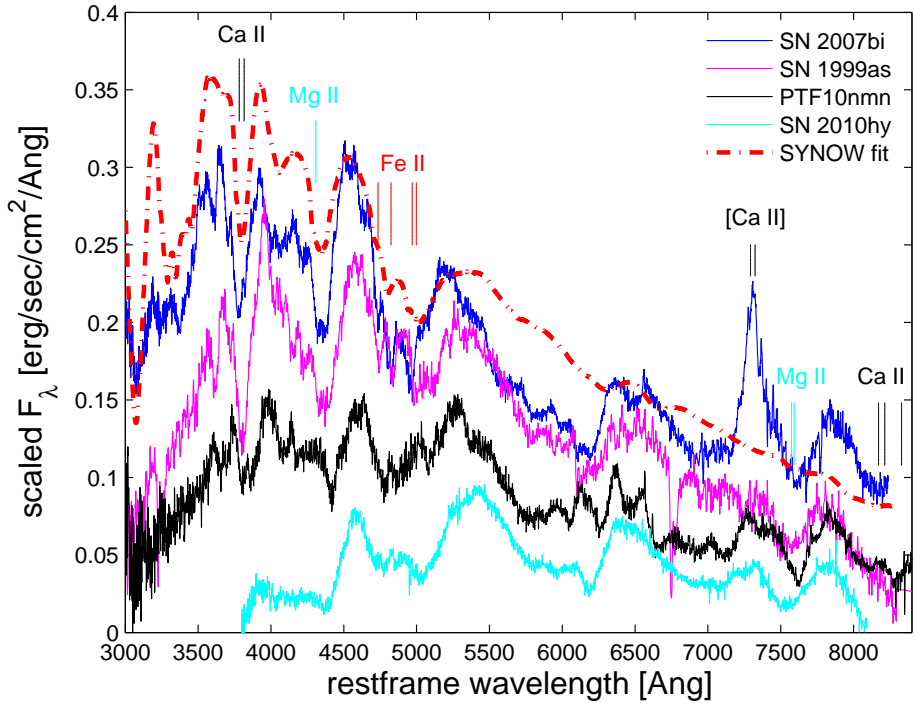}
\caption{Photospheric spectra of SLSN-R events SN 2007bi (blue, from \cite[Gal-Yam et al. 2009]{G+09}), 
SN 1999as (magenta, from \cite[Gal-Yam et al. 2009]{G+09}; Nugent et al. 2012, in preparation),  
PTF10nmn (black, from Yaron et al. 2012, in preparation) and SN 2010hy (cyan; S. B. Cenko, 
private communication); all spectra were obtained close to peak. 
Identification of prominent spectral features as well as a synthetic SYNOW fit (red, 
from \cite[Gal-Yam et al. 2009]{G+09}) are also shown. Figure taken from \cite[Gal-Yam 2012]{G12}.
}
\label{PISNfig}
\end{figure}

SN 1999as (\cite[Knop et al. 1999]{K+99}) was one of the first genuine 
Superluminous Supernovae (SLSN; \cite[Gal-Yam 2012]{G12}) discovered. 
It was initially analyzed by \cite[Hatano et al. (2001)]{H+01}. 
As shown in \cite[Gal-Yam et al. (2009]{G+09}; Fig.~\ref{PISNfig}) 
this object was similar to the likely PISN event SN 2007bi during its photospheric phase
(reaching $-21.4$\,mag absolute at peak).
The analysis of \cite [Hatano et al. (2001)]{H+01} suggests physical 
attributes ($^{56}$Ni mass, kinetic energy, and ejected mass) that are close to, but somewhat
lower than, those of SN 2007bi. Unfortunately, no late-time data have been published for this object, 
so it is impossible to conduct the same analysis carried for SN 2007bi, but the similarities suggest 
this may also have been a PISN. Interestingly, late-time photometry presented for the first time
by K. Nomoto during a presentation in this symposium may argue against this possibility. 

\section{SN 2007bi: the first likely detection of a pair-instability event}

The first well-observed example of a likely PISN was SN 2007bi, discovered by the PTF ``dry run'' experiment. An extensive investigation of this object and its physical nature is presented in \cite[Gal-Yam et al. (2009)]{G+09}. The most prominent physical characteristic of this object
(the prototype of the SLSN-R group), the large $^{56}$Ni mass, is well-measured in this case using both the peak luminosity ($R=-21.3$\,mag) and the cobalt decay tail, followed for $>500$ days. Estimates derived from the observations, as
well as via comparison to other well-studied events (SN 1987A and SN 1998bw) converge on a value of  M$_{^{56}{\rm Ni}}\approx5$\,M$_{\odot}$.
The large amount of radioactive material powers a long-lasting phase of nebular emission, during which the optically thin ejecta are energized by the 
decaying radio nucleides. Analysis of late-time spectra obtained during this phase (\cite[Gal-Yam et al. 2009]{G+09}) provides independent confirmation of the large 
initial $^{56}$Ni mass via detection of strong nebular emission from the large mass of resulting $^{56}$Fe, as well as the integrated emission from
all elements, powered by the remaining $^{56}$Co. 

Estimation of other physical parameters of the event, in particular the total ejected mass (which provides a lower limit on the progenitor star mass), its 
composition, and the kinetic energy it carries, is more complicated. There are no observed signatures of hydrogen in this event (either in the ejecta or
traces of CSM interaction) so the ejecta mass directly constrains the mass of the exploding helium core, which is likely dominated by oxygen and heavier elements. \cite[Gal-Yam et al. (2009)]{G+09} use scaling relations based on the work of \cite[Arnett (1982)]{A82}, as well as comparison of the data to custom light-curve models, and derive an ejecta mass of M\,$\approx100$\,M$_{\odot}$. Analysis of the nebular spectra provides an independent lower limit on the mass of  M\,$>50$\,M$_{\odot}$,
with a composition similar to that expected from theoretical models of massive cores exploding via the pair-instability process.
\cite[Moriya et al. (2010)]{M+10} postulate a lower ejecta mass (M\,$=43$\,M$_{\odot}$); this difference becomes crucial to the controversy about the explosion mechanism of these giant cores (see below). In any case there is no doubt this explosion was produced by an extremely massive star, with the most 
massive exploding heavy-element core we know. The same scaling relations used by \cite[Gal-Yam et al. (2009)]{G+09} 
also indicate extreme values of ejecta kinetic
energy (approaching E$_k=10^{53}$\,erg). Finally, the integrated radiated energy of this event over its very long lifetime is high ($>10^{51}$\,erg).       
   
\section{Additional events}

Recently, the Lick Observatory Supernova Survey (LOSS; \cite[Filippenko et al. 2001]{F+01}) using the 0.75m Katzman Automatic Imaging Telescope (KAIT) discovered the luminous Type Ic SN 2010hy (\cite[Kodros et al. 2010]{K+10}; \cite[Vinko et al. 2010]{V+10}); Following the discovery by KAIT this event was also recovered in PTF data 
(and designated PTF10vwg). It is interesting to note that while the LOSS survey is operating in a targeted mode looking at a list of known galaxies,
by performing image subtraction on the entire KAIT field of view it is effectively running in parallel also an untargeted survey of the background
galaxy population (as noted by \cite[Gal-Yam et al. 2008]{G+08} and \cite[Li et al. 2011]{L+11}). 
It is during this parallel survey that KAIT detected this interesting rare SN,
residing in an anonymous dwarf host. While
final photometry is not yet available for this event, preliminary KAIT and PTF data indicate a peak magnitude of $-21$\,mag or brighter, and it is 
spectroscopically similar to other SLSN-R (Fig.~\ref{PISNfig}) suggesting it is also likely a member of this class.

\begin{figure}[h]
\centering
\includegraphics[width=1\textwidth]{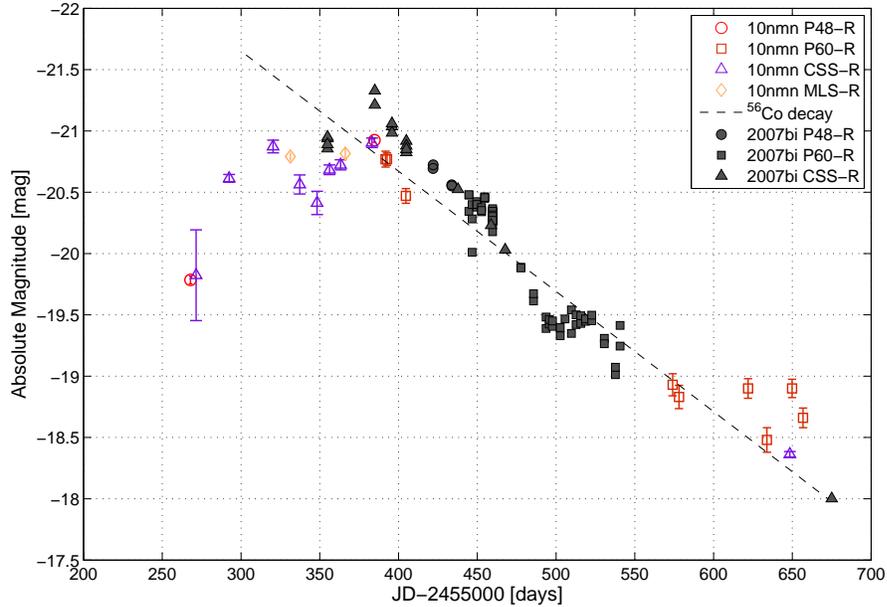}
\caption{Comparison of the light curves of PTF10nmn and SN 2007bi. The luminous
peak and slow decline are similar, indicating a large mass of $^{56}$Ni is powering
these explosions, mixed into a large total mass of ejecta. From Yaron et al. 2012 (in preparation). 
}
\label{Fig10nmn}
\end{figure}

PTF has discovered another likely PISN, PTF10nmn (Yaron et al. 2012, in preparation; Fig.~\ref{Fig10nmn}). The object is similar to 
SN 2007bi both in terms of its light curve (Fig.~\ref{Fig10nmn}) and spectra (Fig.~\ref{PISNfig}). 
Objects of this sub-class are exceedingly rare (this is observationally the rarest class among the SLSN classes; \cite[Gal-Yam 2012]{G12}), 
and thus additional examples are scarce. In addition to this single event from PTF, the Pan-Starrs 1 survey 
may have discovered another similar object at a higher redshift (Kotak et al. 2012, in preparation), while \cite[Cooke et al. (2012)]{C+12}
may have recovered events at even higher redshifts (up to $z\sim4$) in archival SNLS data. 
Assembling a reasonable sample of such events may thus be a time-consuming process. 

\section{The physical properties of SLSN-R and their PISN nature}

Of all classes of super-luminous SNe, this seems to be the best understood. SLSN-R events are powered by large amounts (several M$_{\odot}$) 
of radioactive $^{56}$Ni (hence the suffix ``R''), produced during the explosion of a very massive star. The radioactive decay chain  $^{56}$Ni$\rightarrow^{56}$Co$\rightarrow ^{56}$Fe deposits energy via $\gamma$-ray and positron emission, that is thermalized and converted to optical radiation by the expanding massive ejecta. The luminosity of the peak is broadly proportional to the amount of radioactive $^{56}$Ni, while the late-time decay (which in the most luminous
cases begins immediately after the optical peak) follows the theoretical $^{56}$Co decay rate ($0.0098$\,mag\,day$^{-1}$). The luminosity of this
``cobalt radioactive tail'' can be used to infer an independent estimate of the initial $^{56}$Ni mass.

Considering all available data, it seems there is agreement about the observational properties of this class and their basic interpretation: very massive star explosions that produce large quantities of radioactive $^{56}$Ni. A controversy still exists about the underlying explosion mechanism that leads to
this result, either very massive oxygen cores (M$>50$\,M$_{\odot}$) become unstable to electron-positron pair production and collapse (\cite[Gal-Yam et al. 2009]{G+09}), or else slightly less massive cores (M$<45$\,M$_{\odot}$) evolve all the way till the common iron-core-collapse process occurs (\cite[Moriya et al. 2010]{M+10}). 

\cite[Umeda \& Nomoto (2008)]{UM08} and \cite[Moriya et al. (2010)]{M+10} show that if one considers a 
carbon-oxygen core with a mass of $\sim43$\,M$_{\odot}$ (just below the pair-instability threshold), 
which explodes with an ad-hoc large explosion energy ($>10^{52}$\,erg),
one can produce the required large amounts of nickel (\cite[Umeda \& Nomoto 2008]{UM08}), as well as recover the light curve shape of the
SLSN-R prototype, SN 2007bi (\cite[Moriya et al. 2010]{M+10}). Since both the pair-instability model and the massive core-collapse 
model fit the light curve shape of SN 2007bi equally well; and progenitors of pair-instability explosions have larger cores and thus
larger initial stellar masses, which are, assuming a declining initial mass function, intrinsically more rare, \cite[Yoshida \& Umeda (2011)]{YU11}
favor the core-collapse model. 

\begin{figure}[h]
\centering
\includegraphics[width=1\textwidth]{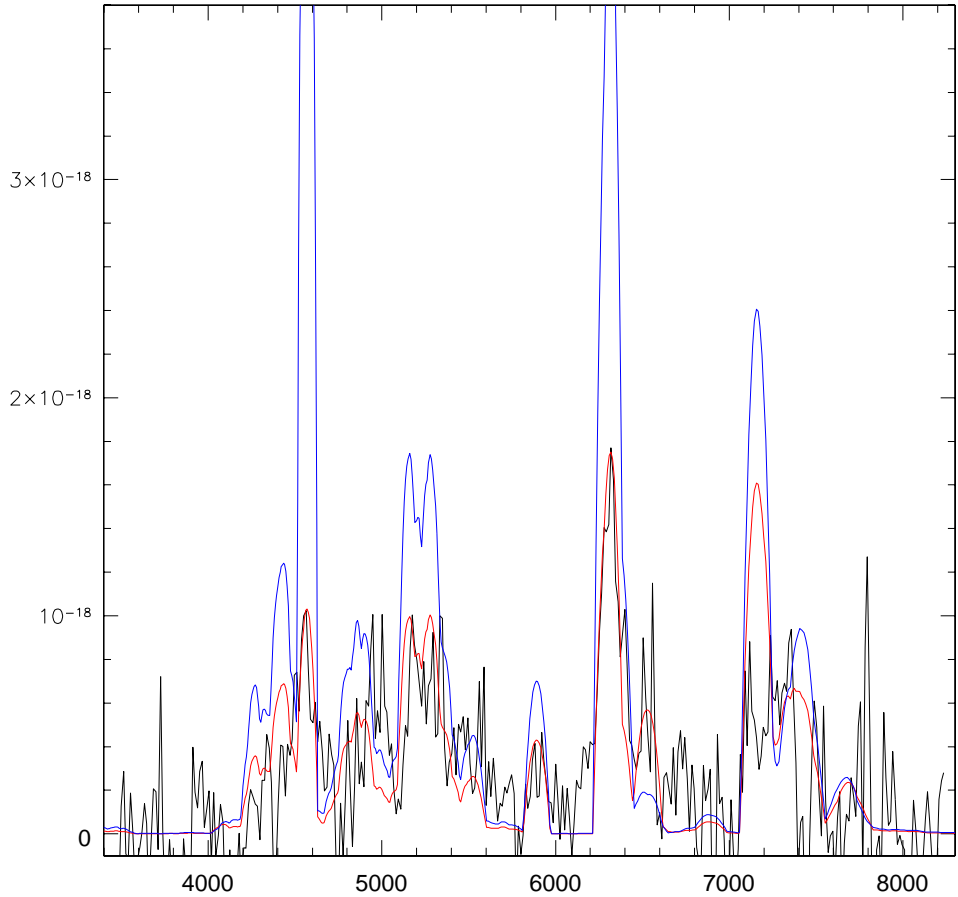}
\caption{A model of the nebular emission expected from ejecta with the composition given by the massive core-collapse model
of \cite[Moriya et al. (2010, blue)]{M+10}, compared to the nebular model from \cite[Gal-Yam et al. 2009]{G+09} (red) 
based on pair-instability event with a composition expected from
the models of \cite[Heger \& Woosley (2002)]{HW02}. The massive core-collapse model (blue) has a similar
amount of radioactive nickel mixed into a smaller total ejecta mass, significantly over-predicting the observed line strengths  
in the nebular spectrum of SN 2007bi (black).
}
\label{sn2007bi-neb}
\end{figure}

The two models (pair instability vs. core-collapse) agree about the nickel mass, but strongly differ in their predictions about
the {\it total} ejected mass. Total heavy-element masses above the $50$\,M$_{\odot}$ threshold would indicate a core that
is bound to become pair unstable, and will rule out the core-collapse model. \cite[Gal-Yam et al. (2009)]{G+09} estimated the total ejected 
heavy-element mass of SN 2007bi in several ways, including modelling of the nebular spectra of this event.  
The core-collapse model of \cite[Moriya et al. 2010]{M+10} does not fit these data (Fig.~\ref{sn2007bi-neb}); this model assumes
a similar amount of radioactive $^{56}$Ni and lower total ejected mass (to avoid the pair-instability) leading to very strong
nebular emission lines that are not consistent with the data. Thus this model is not
viable for this prototypical SLSN-R object, supporting instead a pair instability explosion as originally claimed.
 
It remains to be seen whether the massive core-collapse model does manifest in nature (the prediction would be for
SLSNe showing large amounts of radioactive nickel but relatively small amounts of total ejecta). As a final note, it should be
stressed that while the stellar evolution models considered by \cite[Yoshida \& Umeda (2011)]{YU11} require stars with exceedingly large initial 
masses ($>310$\,M$_{\odot}$) to form pair-unstable cores at the moderate metallicity indicated for SN 2007bi (\cite[Young et al. 2010)]{Y+10}, 
alternative models (\cite[Langer et al. 2007]{L+07}) predict that stars with much lower initial masses ($150-250$\,M$_{\odot}$) explode as pair-instability SNe at 
SMC- or LMC-like metallicities, though they may have to be tweaked to
explain the lack of hydrogen in observed SLSN-R spectra. 

\begin{figure}[h]
\centering
\includegraphics[width=1\textwidth]{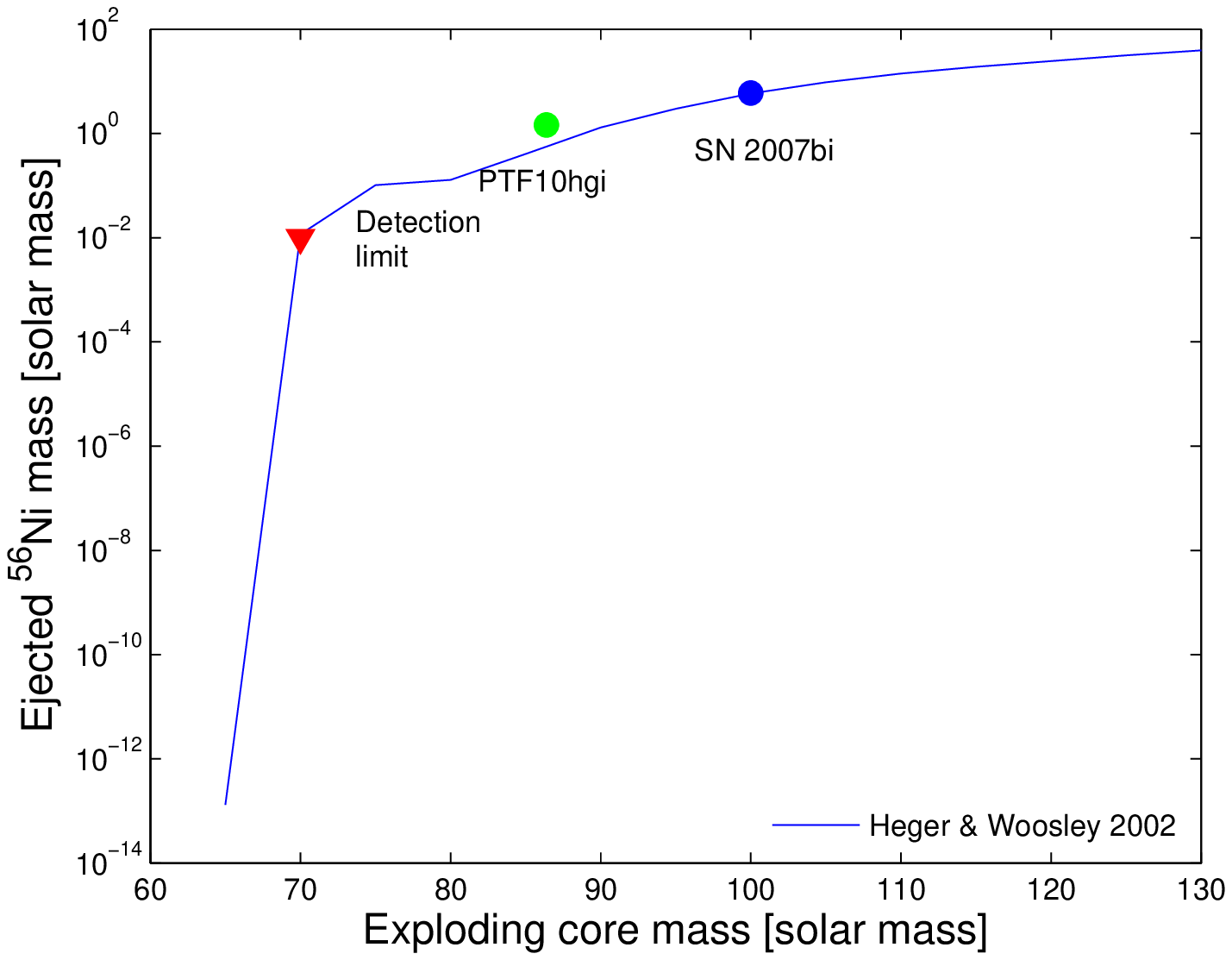}
\caption{The relation between the synthesized $^{56}$Ni mass and the He-core mass (which should roughly equal the 
total ejected mass for Type I events) based on the (non-rotating and non-magnetic) models of 
\cite[Heger \& Woosley (2002)]{HW02}. Superposed are the data points for SN 2007bi (\cite[Gal-Yam et al. 2009]{G+09}) 
and PTF10hgi (previously unpublished). 
We also mark the approximate upper limit on object detectability (M$_{^{56}{\rm Ni}}=0.01$\,M$_{\odot}$) below which these 
events are expected to be too faint to be discovered anywhere except for the nearest galaxies, where the expected rate of these
events is probably prohibitively low.   
}
\label{PISN-stats}
\end{figure}

Assuming, for the sake of the current discussion, that observed 
SLSN-R explosions do arise from the pair instability, a clear prediction of the relevant theoretical
models (e.g., \cite[Heger \& Woosley 2002]{HW02}, \cite[Waldman 2008]{W08}) 
is that for each luminous, $^{56}$Ni-rich explosion (from a core around $100$\,M$_{\odot}$) there
would be numerous less luminous events with smaller $^{56}$Ni masses but large ejecta masses (M$>50$\,M$_{\odot}$). These should manifest
as events with very slow light curves (long rise and decay times) and yet moderate or even low peak luminosities (Fig.~\ref{PISN-stats}). 

\section{Host galaxies}

\cite[Young et al. (2010)]{Y+10} present a detailed study of the host galaxy of SN 2007bi.
They find the host is a dwarf galaxy (with luminosity similar to that of the SMC), with relatively low 
metallicity ($Z\approx Z_{\odot}/3$) - somewhere between those of the LMC and SMC. 
So, while the progenitor star of this explosion probably had sub-solar metal content, there
is no evidence that it had very low metallicity. The host galaxy
of SN 1999as is more luminous (and thus likely more metal-rich) than that of SN 2007bi, 
but still fainter than typical giant galaxies like the Milky Way (\cite[Neill et al. 2011]{N+11}), 
while the host of PTF10nmn seems to be as faint or fainter than that of SN 2007bi (Yaron et al. 2012,
in preparation). It thus seems this class of objects typically explode in dwarf galaxies. This
is likely yet another aspect of the difference between the population of massive star 
explosions in observed in dwarf galaxies compared to giant hosts (\cite[Arcavi et al. 2010]{A+11}). 

\section{Rates}

An estimate of the rate of SLSN-R can be derive from a 
rough estimate of the rate of SLSN-I provided by \cite[Quimby et al. (2011)]{Q+11} based on statistics of events
detected by the Texas Supernova Survey (TSS), which,
normalizing the rate of SLSN-I at $z\approx 0.3$ relative to that of SNe Ia, yields a SLSN-I rate of $\sim10^{-8}$\,Mpc$^{-3}$\,y$^{-1}$.  
Both the reported discovery statistics as well as unpublished PTF counts suggest that SLSN-R are rarer by about a factor of five, 
correcting for their slightly lower peak luminosities. 
This rate is substantially lower than the rates of core-collapse SNe ($\sim10^{-4}$\,Mpc$^{-3}$\,y$^{-1}$), and is also well
below those of rare sub-classes like broad-line SNe Ic (``hypernovae''; $\sim10^{-5}$\,Mpc$^{-3}$\,y$^{-1}$) or long Gamma-Ray 
Bursts ($>10^{-7}$\,Mpc$^{-3}$\,y$^{-1}$; \cite[Podsiadlowski et al. 2004]{P+04}; \cite[Guetta \& Della Valle 2007]{GD07}).
Interestingly, in is comparable to the rate recently predicted by \cite[Pan, Loeb \& Kasen (2012)]{P+12}. 
I believe this suggests that SLSN-R are indeed the rarest type of explosions studied so far, and quite possibly arise from 
stars that are at the very top of the IMF.

%\begin{discussion}

%\discuss{Massey}{I'm wondering if you have considered the expected intrinsic dispersion in absolute
%magnitude of WRs --- if you consider the (large) mass range that becomes an
%early WN or late WC according to the evolutionary models, wouldn't you expect a large
%dispersion in M$_v$?}

%\end{discussion}

\end{document}